\documentclass[reprint,amsmath,amssymb,aip,superscriptaddress,longbibliography]{revtex4-1}
\pdfoutput=1

\usepackage{hyperref,url}
\usepackage{amsmath}
\usepackage{amsfonts}
\usepackage{braket}
\usepackage{siunitx}
\usepackage[capitalise]{cleveref}
\usepackage{placeins}
\hypersetup{breaklinks,colorlinks=true,linkcolor=blue,citecolor=blue,urlcolor=blue,filecolor=blue}
\usepackage{graphicx}
\usepackage{dcolumn}
\usepackage{tikz}
\usepackage[inline]{enumitem}
\usetikzlibrary{arrows.meta, positioning, shapes, shadows, trees}
\tikzset{
    basic/.style  = {draw, text width=5cm, drop shadow, font=\sffamily, rectangle},
    root/.style   = {basic, rounded corners=2pt, thin, align=center,
                     fill=green!60},
    level 2/.style = {basic, rounded corners=6pt, thin,align=center, fill=green!30,
                     text width=9.5em},
    level 3/.style = {basic, thin, align=left, fill=pink!60, text width=6.5em}
}
\usepackage{algorithm}
\usepackage{mhchem}
\usepackage{algpseudocode}
\usepackage{listings}
\usepackage{xcolor}

\definecolor{codegreen}{rgb}{0,0.6,0}
\definecolor{codegray}{rgb}{0.5,0.5,0.5}
\definecolor{codepurple}{rgb}{0.58,0,0.82}
\definecolor{backcolour}{rgb}{0.95,0.95,0.92}

\lstdefinestyle{mypython}{
    language=Python,
    backgroundcolor=\color{backcolour},   
    commentstyle=\color{codegray},
    keywordstyle=\color{magenta},
    numberstyle=\tiny\color{codegray},
    stringstyle=\color{codegreen},
    basicstyle=\ttfamily\footnotesize,
    breakatwhitespace=false,         
    breaklines=true,                 
    captionpos=b,                    
    keepspaces=true,                 
    numbers=left,                    
    morekeywords=[1]{einsum,dot,inv,empty,copy,reshape,zeros,shape,index,to_csv,run,kernel},
    numbersep=5pt,                  
    showspaces=false,                
    showstringspaces=false,
    showtabs=false,                  
    tabsize=2
}

\lstnewenvironment{python}[1][]{\lstset{style=mypython,#1}}{}

\Crefname{lstlisting}{Listing}{Listings}

\newcommand\mat\mathbf

\newcommand{\trexio}{\textsc{TrexIO}}

\usepackage{ulem}

\newcommand{\ByteDance}{\affiliation{ByteDance Research, Zhonghang Plaza, No. 43,
North 3rd Ring West Road, Haidian District, Beijing 100089, China}}
\newcommand{\ByteDanceUS}{\affiliation{ByteDance Research, San Jose, CA 95110, US.}}

\begin{document}

\author {Yifei Huang}
\email{huangyifei.426@bytedance.com}
\ByteDance
\author {Zhen Guo}
\ByteDance
\author {Hung Q. Pham}
\ByteDanceUS
\author {Dingshun Lv}
\email{lvdingshun@bytedance.com}
\ByteDance

\title{GPU-accelerated Auxiliary-field quantum Monte Carlo with multi-Slater determinant trial states
 }
\begin{abstract}
The accuracy of phaseless auxiliary-field quantum Monte Carlo~(ph-AFQMC) can be systematically improved with better trial states. Using multi-Slater determinant trial states, ph-AFQMC has the potential to faithfully treat strongly correlated systems, while balancing the static and dynamical correlations on an equal footing.
This preprint presents an implementation and application of graphics processing unit-accelerated ph-AFQMC, for multi-Slater determinant trial wavefunctions (GPU-accelerated MSD-AFQMC), to enable efficient simulation of large-scale, strongly correlated systems. This approach allows for nearly-exact computation of ground state energies in multi-reference systems. Our GPU-accelerated MSD-AFQMC is implemented in the open-source code \texttt{ipie}, a Python-based AFQMC package [\textit{J. Chem. Theory Comput.}, 2022, 19(1): 109-121]. We benchmark the performance of the GPU code on transition-metal clusters like [Cu$_2$O$_2$]$^{2+}$ and [Fe$_2$S$_2$(SCH$_3$)]$^{2-}$. The GPU code achieves at least sixfold speedup in both cases, comparing the timings of a single A100 GPU to that of a 32-CPU node. For [Fe$_2$S$_2$(SCH$_3$)]$^{2-}$, we demonstrate that our GPU MSD-AFQMC can recover the dynamical correlation necessary for chemical accuracy with an MSD trial, despite the large number of determinants required~($>10^5$). Our work significantly enhances the efficiency of MSD-AFQMC calculations for large, strongly correlated molecules by utilizing GPUs, offering a promising path for exploring the electronic structure of transition metal complexes.
\end{abstract}
\maketitle

\section{Introduction}
Auxiliary-field quantum Monte Carlo~(AFQMC)~\cite{zhang_constrained_1995,Zhang2003Apr} has gained increasing popularity in quantum chemistry~\cite{lee_twenty_2022,motta_ab_2018}
and is also recognized as a powerful approach within the realm of quantum algorithms~\cite{huggins_unbiasing_2022,wan2023matchgate,amsler2023classical,kiser2023classical,Huang2024Apr}.
AFQMC with phaseless approximation~(ph-AFQMC) imposes constraints on the evolution paths of walkers, thereby suppressing the sign problem while introducing biases associated with the trial states. As the quality of the trial states improves, one can systematically reduce the biases in AFQMC calculations. The gold standard method of quantum chemistry, coupled cluster singles and doubles with perturbative triples (CCSD(T)), is known to struggle in achieving chemical accuracy for \textit{ab initio} systems with multi-reference characteristics, including bond dissociation and transition metal complexes. The accuracy of AFQMC with single-determinant trial states~(SD-AFQMC) performs even worse than the (CCSD(T)) results for these systems~\cite{lee_twenty_2022}. By employing more sophisticated trial states with multi-Slater determinants~(MSD-AFQMC), the accuracy of AFQMC has been shown to approach the exact results for several benchmark multi-reference systems. More specifically, the accuracy of MSD-AFQMC is on par with the multi-reference configuration
interaction with Davidson correction~(MRCI+Q) method, while having a better scaling and can be extended to larger systems~\cite{lee_twenty_2022}. These benchmarks positions MSD-AFQMC as one of the front-runners for large-scale calculations of multi-reference systems, where static and dynamic correlations need to addressed on an equal footing~\cite{shee2023potentially}.

\texttt{ipie} is a Python-based AFQMC code that was optimized for high-performance computing on central processing units (CPUs), with a focus on simplicity and speed. It integrates \texttt{Numba}'s JIT compilation to enhance the computational efficiency of specific kernels and uses MPI parallelism for effective distributed computing. While engineering efforts were directed towards GPUs for single-Slater determinant trials (SD-AFQMC), the MSD-AFQMC code has only been implemented on general-purpose CPUs due to the complicated excitations in multi-determinant trial states. This underscores the pressing need for MSD-GPU development to advance large-scale calculations for transition metal systems. In this preprint, we present our GPU implementation and benchmarking of MSD-AFQMC as an important extended feature of \texttt{ipie}. We implemented Nvidia Compute Unified Device Architecture (CUDA) kernel functions with a \texttt{cupy} interface to facilitate parallel treatments of the different excitations in the MSD trials within \texttt{ipie}.

\section{Theory}\label{sec:afqmc}
\subsection{Phaseless AFQMC}

AFQMC is a projector quantum Monte Carlo method that realises the imaginary time evolution~(ITE):
\begin{equation}
    \left|\Psi_0\right\rangle \propto \lim _{\tau \rightarrow \infty} \exp (-\tau \hat{H})\left|\Phi_0\right\rangle=\lim _{n\rightarrow \infty}(\exp 
 (-\tau \hat{H}/n))^n|\Phi_0\rangle, \label{eq:afqmc}
\end{equation}
where $\left|\Psi_0\right\rangle$ is the ground state wavefunction, and $\left|\Phi_0\right\rangle$ is an initial state that have a non-zero overlap with the ground state.
We pivot our GPU engineering to \textit{ab initio} Hamiltonian, which, in second quantization, is given by
\begin{equation}
\hat{H}=\sum_{p,q=1}^N h_{p q} \hat{a}_p^{\dagger} \hat{a}_q+\frac{1}{2} \sum_{p,q,r,s=1}^N g_{psqr} \hat{a}_p^{\dagger} \hat{a}_q^{\dagger} \hat{a}_r \hat{a}_s. \label{eq:qcham}
\end{equation}
where $h_{pq}$ and $g_{pqrs}$ are the 1-electron and 2-electron repulsion integrals~(ERI). This can be rewritten as 
\begin{equation}
\hat{H}=\sum_{p,q=1}^N h'_{p q} \hat{a}_p^{\dagger} \hat{a}_q+\frac{1}{2}\sum_{\gamma=1}^{N_\gamma}\left( \sum_{p q} L_{p q}^\gamma \hat{a}_{p }^{\dagger} \hat{a}_{q}\right)^2
\end{equation}
using the Cholesky decomposition to the two-electron repulsion integral~(ERI). Here, 
we have the 
\begin{equation}
h'_{pq} = h_{pq} -\frac{1}{2}\sum_r (pr|rq)
\end{equation}
Applying the Hubbard--Stratonovich transformation to the two-body term\cite{hubbard1959calculation,Stratonovich}, the effective propagator contains only one-body operators,
\begin{equation}
    e^{-\Delta \tau \hat{H}} = \int \mathrm{d} \mathbf{x} \ p(\mathbf{x}) \hat{B}(\mathbf{x})+\mathcal{O}\left(\Delta \tau^2\right),
\end{equation}
where $p(\mathbf x)$ is the standard Gaussian distribution, $\textbf{x}=(x_1, x_2, \cdots, x_{N_{\gamma}})$ are the auxiliary fields, and $\hat{B}$ is a propagator with only one-body terms on its exponents. Here, we have introduced the first-order Trotter decomposition, and therefore, the $\mathcal{O}\left(\Delta \tau^2\right)$ Trotter error term.

In AFQMC, each walker is a single Slater determinant, which samples auxiliary fields $\mathbf{x}_i $ and becomes another single Slater determinant due to the Thouless' theorem: $B(\mathbf{x}_i)\ket{\phi_i(\tau)}=\ket{\phi_i(\tau+\Delta\tau)}$. Each walker represents a statistical sample of the global wavefunction at imaginary time $\tau$, written as 
\begin{equation}
|\Psi(\tau)\rangle=\sum_i^{N_\textrm{w}} w_i(\tau) \frac{\left|\phi_i(\tau)\right\rangle}{\left\langle\Psi_{T}|\phi_i(\tau)\right\rangle},
\end{equation}
where $w_i(\tau)$ is the weight, $|\phi_i(\tau)\rangle$ is the wavefunction of the $i$-th walker at time $\tau$ and $|\Psi_{T}\rangle$ is the trial wavefunction used for importance sampling.

We dynamically shift the distribution of auxiliary fields using the force bias $\overline{\mathbf{x}}_i$ defined by
\begin{equation}
    \overline{x}_{\gamma}(\Delta \tau, \tau)=-\sqrt{\Delta \tau} \frac{\left\langle\Psi_{T}\left|\sum_{p q} L_{p q}^\gamma \hat{a}_{p }^{\dagger} \hat{a}_{q}\right| \psi_i(\tau)\right\rangle}{\left\langle\Psi_{T}|\psi_i(\tau)\right\rangle}.\label{eq:fb}
\end{equation}

At each imaginary time step, the overlap ratio of the $i$-th walker is
\begin{equation}
    S_i(\tau, \Delta \tau)=\frac{\left\langle\Psi_{\textrm{T}}\left|\hat{B}\left(\Delta \tau, \mathbf{x}_i-\overline{\mathbf{x}}_i\right)\right| \psi_i(\tau)\right\rangle}{\left\langle\Psi_{\textrm{T}}|\psi_i(\tau)\right\rangle}, \label{eq:ovlp_ratio}
\end{equation}
with the phase of the overlap as
\begin{equation}
    \theta_i(\tau)=\arg \left[S_i(\tau, \Delta \tau)\right]. \label{phase}
\end{equation}
Hence, we can define the importance function for the update of the weights $w_i(\tau)$s:
\begin{equation}
    I\left(\mathbf{x}_i, \overline{\mathbf{x}}_i, \tau, \Delta \tau\right)=S_i(\tau, \Delta \tau) e^{\mathbf{x}_i \cdot \overline{\mathbf{x}}_i-\overline{\mathbf{x}}_i \cdot \overline{\mathbf{x}}_i / 2},
\end{equation}

The phase factor in Eq.~\ref{phase} can range from 0 to $2\pi$ and therefore causes the phase/sign problem. In order to control this phase/sign problem, the phaseless approximation~\cite{Zhang2003Apr} is applied during the update of the weights $w_i(\tau)$:
\begin{equation}
w_i(\tau+\Delta \tau)=I_{\mathrm{ph}}\left(\mathbf{x}_i, \overline{\mathbf{x}}_i, \tau, \Delta \tau\right) \times w_i(\tau).  \label{eq:weight_update}
\end{equation}
where $I_{ph}$ is given by
\begin{equation}
    I_{\mathrm{ph}}\left(\mathbf{x}_i, \overline{\mathbf{x}}_i, \tau, \Delta \tau\right)=\left|I\left(\mathbf{x}_i, \overline{\mathbf{x}}_i, \tau, \Delta \tau\right)\right| \times \max \left[0, \cos \left(\theta_i(\tau)\right)\right],
\end{equation}
which remains real and positive throughout the propagation.

When the trial wavefunction is a single Slater determinant, the overlaps at each time step is simply the overlap between two Slater determinants:
\begin{equation}
\langle \Psi_{T}|{\phi_i}\rangle=\text{det}\left( \Psi_{T}^{\dagger}\phi_i\right)
\end{equation}
where $\Psi_{T}$ and $\phi_i$ are the matrix representation of the two Slater determinants. The Green's function, i.e., the one-body reduced density matrix~(1-RDM) in a mixed form is computed as follows:
\begin{equation}
G_{ij}=\frac{\bra{\Psi_{T}}a_i^{\dagger}a_j\ket{\phi_i}}{\langle{\Psi_{T}}|{\phi_i}\rangle}=\left[\phi_i\left(\Psi_{T}^{\dagger}\phi_i\right]^{-1}\Psi_{T}^{\dagger}\right]
\end{equation}
The two-body RDMs~(2-RDMs) can be obtained from the 1-RDM using the generalized Wicks theorem~\cite{wick1950evaluation,balian1969nonunitary}. The mixed energy estimation can therefore, be calculated by the 1-RDMs and 2-RDMs.

For SD-AFQMC, the only calculations needed are matrix multiplications, as shown above. This makes it fairly straightforward for GPU implementation, with the $N_w$ dimension also batched for parallelization.

\subsection{Multi-Slater determinant AFQMC}

The trial wavefunction in this case is written as a linear combination of different Slater determinants:
\begin{equation}
\ket{\Psi_{T}}=\sum_{j=1}^{N_d}c_j\ket{\phi_j},
\end{equation}
where $N_d$ is the number of determinants in the trial. The Slater determinants can be either a configuration interaction~(CI) expansion, where the determinants are orthogonal to each other under the same set of orbitals, or non-orthogonal under different sets of orbitals~\cite{LandinezBorda2019Feb}. In this preprint, we only work with the former case where each determinant is represented as a configuration.

The naive approach to implementing MSD-AFQMC involves performing SD-AFQMC for each individual determinant in the trial wavefunction. Although this makes GPU parallelization straightforward, it is inefficient because the orbitals used in the CI expansion are the same across determinants. Starting from scratch for each determinant results in linear time scaling with respect to $N_d$. Moreover, fully parallelizing this on GPUs would lead to a linear increase in memory usage with $N_d$, which is not scalable. In contrast, the \texttt{ipie} implementation of MSD-AFQMC achieves sub-linear scaling in $N_d$ by reusing orbital information~\cite{shi2021some}. This advanced implementation demonstrates that the time required for MSD-AFQMC with 2 determinants is nearly the same as with 1000 determinants. However, this efficiency complicates GPU implementation, as the trial wavefunction must be managed carefully to optimize memory usage.


Meanwhile, \texttt{ipie} integrates external quantum chemistry packages to facilitate a convenient workflow for MSD-AFQMC. 
For more complicated trial wavefunctions, such as multiple Slater determinant~(MSD) trials derived from selected configuration interaction~(SCI), interfaces with \texttt{PySCF}~\cite{sun2018pyscf}~(shciscf interface), \texttt{Dice}~\cite{holmes2016heat,sharma2017semistochastic}, and \trexio{}\cite{trexio} are available and easy to use. In this preprint, we use Dice to perform the SCI calculations. 
 

\section{GPU-accelerated MSD-AFQMC and timing benchmarks}\label{sec:msd_gpu}

ph-AFQMC with MSD trials is useful for multi-reference systems where free-projection is not feasible due to the sign problem. 
In this main section, we report our GPU-accelerated MSD-AFQMC workflow based on Wick's theorem~\cite{mahajan2022selected} .
We used \texttt{cupy.einsum} as implemented through the \texttt{cuTENSOR} library to accelerate matrix multiplications. Furthermore, we implemented Wick's algorithm entirely on the GPU with CUDA kernels. These kernels substantially accelerated computational hotspots in MSD-AFQMC, namely calculations of the Green's function, overlap, and local energy. One only need minimal changes to the \texttt{ipie} multi-determinant example to make use of the GPU code:

\begin{lstlisting}[language=Python, caption=Code snippet for MSD calculation with a single GPU., label={lst:from_pyscf}]
%env CUPY_ACCELERATORS=cutensor # for notebook, for .py you can set this in terminal

import cupy

from ipie.config import config

config.update_option("use_gpu", True)

import h5py
import numpy
from pyscf import fci, gto, mcscf, scf

from ipie.hamiltonians.generic import Generic as HamGeneric
from ipie.qmc.afqmc import AFQMC
from ipie.systems.generic import Generic
from ipie.trial_wavefunction.particle_hole import ParticleHole
from ipie.utils.from_pyscf import gen_ipie_input_from_pyscf_chk

nocca = 4
noccb = 2

mol = gto.M(
atom=[("N", 0, 0, 0), ("N", (0, 0, 3.0))],
basis="ccpvdz",
verbose=3,
spin=nocca - noccb,
unit="Bohr",
)
mf = scf.RHF(mol)
mf.chkfile = "scf.chk"
ehf = mf.kernel()
M = 6
N = 6
mc = mcscf.CASSCF(mf, M, N)
mc.chkfile = "scf.chk"
e_tot, e_cas, fcivec, mo, mo_energy = mc.kernel()
coeff, occa, occb = zip(
*fci.addons.large_ci(fcivec, M, (nocca, noccb), tol=1e-8, return_strs=False)
)

with h5py.File("scf.chk", "r+") as fh5:
    fh5["mcscf/ci_coeffs"] = coeff
    fh5["mcscf/occs_alpha"] = occa
    fh5["mcscf/occs_beta"] = occb

gen_ipie_input_from_pyscf_chk("scf.chk", mcscf=True)
mol_nelec = [8, 6]

with h5py.File("hamiltonian.h5") as fa:
chol = fa["LXmn"][()]
h1e = fa["hcore"][()]
e0 = fa["e0"][()]

num_basis = chol.shape[1]
system = Generic(nelec=mol_nelec)

num_chol = chol.shape[0]
ham = HamGeneric(
numpy.array([h1e, h1e]),
chol.transpose((1, 2, 0)).reshape((num_basis * num_basis, num_chol)),
e0,
)

with h5py.File("wavefunction.h5", "r") as fh5:
    coeff = fh5["ci_coeffs"][:]
    occa = fh5["occ_alpha"][:]
    occb = fh5["occ_beta"][:]
wavefunction = (coeff, occa, occb)
trial = ParticleHole(
    wavefunction,
    mol_nelec,
    num_basis,
    num_dets_for_props=len(wavefunction[0]),
    verbose=True,
)
trial.compute_trial_energy = True
trial.build()
trial.half_rotate(ham)

afqmc_msd = AFQMC.build(
    mol_nelec,
    ham,
    trial,
    num_walkers=10,
    num_steps_per_block=25,
    num_blocks=10,
    timestep=0.005,
    stabilize_freq=5,
    seed=96264512,
    pop_control_freq=5,
    verbose=True,
)
afqmc_msd.run()
afqmc_msd.finalise(verbose=True)
\end{lstlisting}
Here one simply needs to set the \texttt{cutensor} environment variable, use the GPU config and make sure the \texttt{ParticleHole} class is used for the trial wavefunction.

\subsection{Implementation}

In MSD-AFQMC, the trial wavefunction comprises various configurations arising from different excitations relative to the reference configuration, complicating GPU parallelization. 

In \texttt{ipie}, the Wick's theorem calculations are organized into four functional sets: \texttt{get\_dets}, \texttt{reduce\_CI}, \texttt{fill\_os}, and \texttt{get\_ss}, where \texttt{ss} and \texttt{os} refer to same-spin and opposite-spin, respectively. Each set handles single, double, triple, and n-fold excitation components. Specifically, \texttt{get\_dets} aids in overlap calculations, \texttt{reduce\_CI} is crucial for evaluating Green's functions, and the last two sets, \texttt{fill\_os} and \texttt{get\_ss}, contribute to the mixed energy estimator.

Our implementation primarily involves element-wise and reduction operations, with meticulous management of various dimensions to balance time and memory usage. Readers interested in more detailed information about the implementation are encouraged to consult the Appendix.

\subsection{Benchmark results}

We benchmark our MSD-AFQMC GPU implementation against the previous CPU version using the [Cu$_2$O$_2$]$^{2+}$ system with the structural parameter $f=0$!\cite{mahajan2021taming} as described in the earlier \texttt{ipie} release paper~\cite{malone_ipie_2023}. We adopt the same BS1 basis used in \citenum{mahajan2022selected} and the correlation space contains (32e, 108o) after a frozen core. To ensure a direct comparison, we employ a total of 10 walkers, the same as the configuration used in the reference benchmark~\cite{malone_ipie_2023}. Additionally, we conduct tests with 640 walkers to provide a more realistic evaluation. Consistent with earlier benchmarks, we define a block as 25 propagation steps, incorporating one local energy estimation and executing population control every 5 steps.

\begin{figure}
    \centering
    \includegraphics[width=0.45\textwidth]{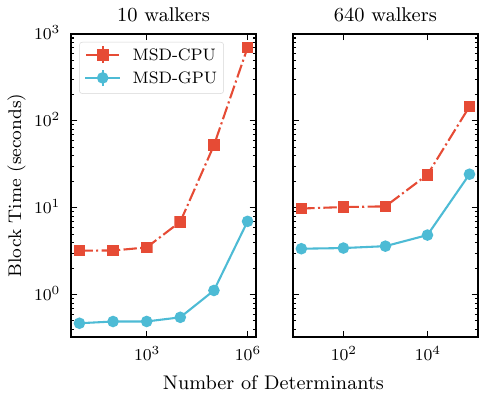}
    \caption{(a) Comparison of time per MSD-AFQMC block with 10 walkers on a single CPU core and a single A100 GPU. The system we considered was [Cu$_2$O$_2$]$^{2+}$ using the BS1 basis~\cite{mahajan2022selected}, which was also a CPU time benchmark example in the original ipie release~\cite{malone_ipie_2023}. (b) Time per block with 640 walkers on a single A100 against 32 CPU cores.}
    \label{fig:msd_gpu}
\end{figure}

With 10 walkers, we find that our GPU implementation on a single NVIDIA A100 card is six times faster than the CPU version on a single CPU core when the number of determinants in the trial is less than $10^3$. The speedup increases to tenfold with $10^4$ determinants and about 100-fold with $10^6$ determinants (see Fig.~\ref{fig:msd_gpu}(a)). This is expected as the GPU utility rate increases with more determinants in the trial. This scenario is particularly relevant when we have many determinants but a small number of walkers (i.e., 10) in each MPI process.

With 640 walkers, we compare the performance of our GPU code on a single A100 card against our CPU implementation on 32 CPU cores, reflecting typical production-level AFQMC calculations. As depicted in Fig.~\ref{fig:msd_gpu}
(b), we observe a fourfold speedup with fewer than $10^2$ determinants, which increases to approximately sixfold as the number of determinants grows. To tackle a larger number of determinants within the memory constraints, we partition the determinants into chunks and compute each chunk sequentially. This partition introduces overhead that primarily impacts the energy estimator. For instance, with $10^3$ determinants, the estimator accounts for less than 50\% of the total block time (0.38s on average out of 0.91s); with $10^6$ determinants, this ratio increases to 75\% (37s out of 49s). A different approach for computing local energy may be required for systems that necessitate numerous determinants in the MSD trial.

\begin{figure}
    \centering
    \includegraphics[width=0.4\textwidth]{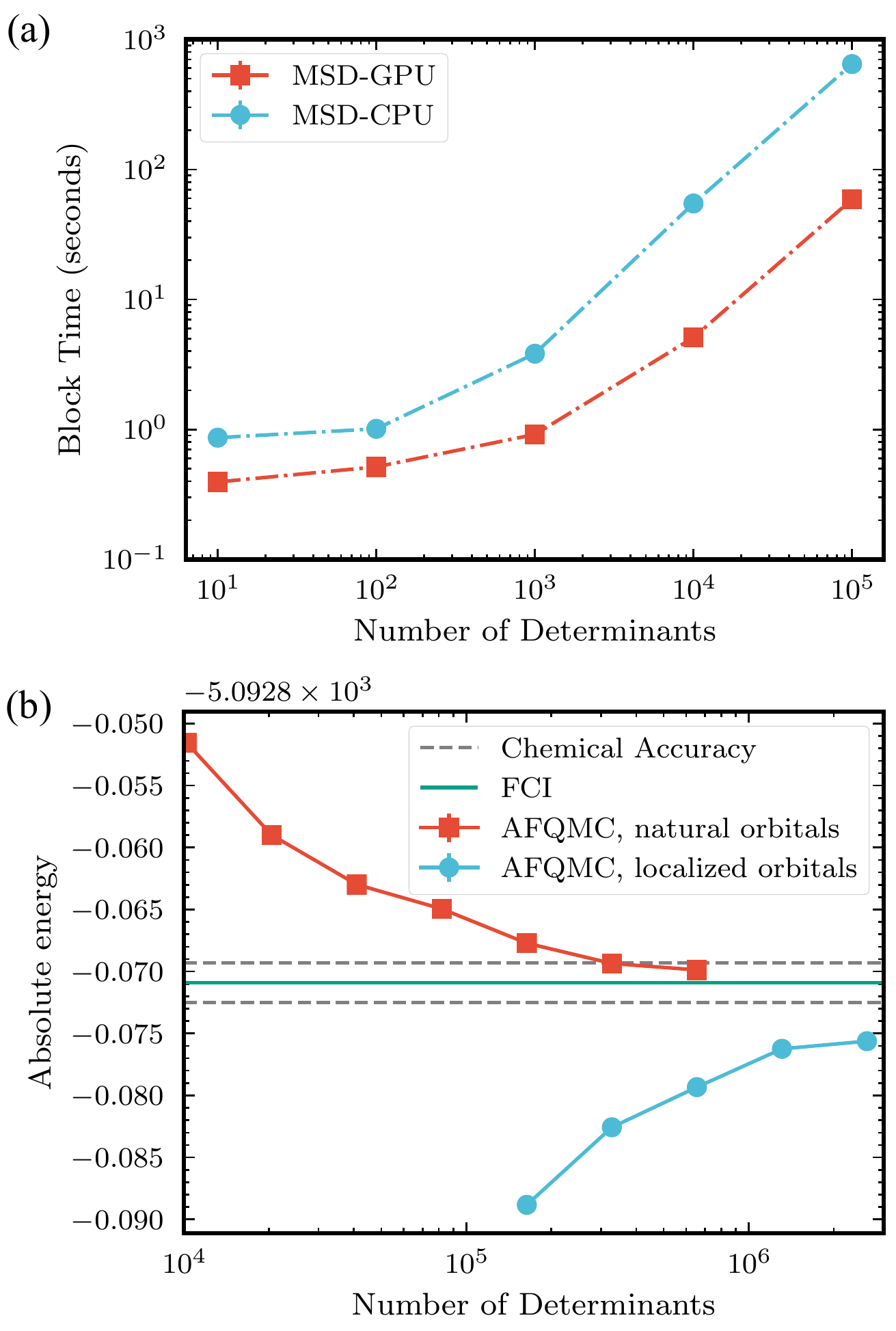}
    \caption{Timing and absolute energy benchmarks on the [Fe$_2$S$_2$(SCH$_3$)]$^{2-}$ cluster. (a) Comparison of time per MSD-AFQMC block with 640 walkers on 32 CPU cores and a single A100 GPU. (b) Absolute energies derived using localized atomic orbitals and natural orbitals. The full configuration interaction result is used as reference.}
    \label{fig:msd_gpu_fes}
\end{figure}

The iron-sulfur clusters is known for its notoriously strong correlation due to the large number of d-orbital electrons in iron and the double exchange mechanism together with the sulfur p-orbital electrons~\cite{sharma2014low,li2019electronic}. Here we performed timing and absolute energy benchmarks on the (20o, 30e) active space of the [Fe$_2$S$_2$(SCH$_3$)]$^{2-}$ cluster. The active space and integrals were obtained from Refs.~\citenum{sharma2014low} and~\citenum{li2017spin}. In Fig.~\ref{fig:msd_gpu_fes}(a), one can see that in contrast to the 108-orbital calculation of [Cu$_2$O$_2$]$^{2+}$, we only have a twofold speedup using a single A100 card compared to 32 CPUs. This speedup increases to almost tenfold when the number of determinants exceeds $10^4$.

For absolute energies, we compute the reference energy from full configuration interaction~(FCI) and use two sets of orbitals for this benchmark. The first set consists of natural orbitals obtained by diagonalizing the 1-electron reduced density matrix (1-RDM) from a high-level wavefunction~\cite{lee2023evaluating}.
Performing MSD-AFQMC calculations based on these orbitals, we can reach within chemical accuracy from the FCI result with around $3\times10^5$ determinants in the trial. On the other hand, reaching chemical accuracy is more challenging when we use localized orbitals from atomic orbitals. As shown in Fig.~\ref{fig:msd_gpu_fes}(b), the difference from the reference value is still 4mHa with more than $2.5\times10^6$ determinants. These observations indicate the importance of choosing a proper set of orbitals for calculations on strongly correlated systems. Here we cheated a little bit using natural orbitals resulted from a near-exact DMRG calculation, which is not viable for larger systems. However, we remark that we can always perform calculations with lower bond dimensions~(DMRG) or larger thresholds~(SCI) and obtain the natural orbitals thereafter. Another interesting point about the mixed energy estimator is that the energies derived using localized orbitals with a limited number of determinants are lower than the FCI energy and approach upwards toward the FCI reference.

\section{Conclusions and outlooks}\label{sec:conclusion}

In this preprint, we present an efficient and user-friendly GPU implementation of MSD-AFQMC, which we believe will serve as a valuable tool for both methodology developers and electronic structure practitioners aiming to perform large-scale calculations on multi-reference systems. These systems require a balanced treatment of both static and dynamic correlations for accurate descriptions. On the other hand, as one can view AFQMC as a tool to make up for dynamical correlation beyond the active space, our [Fe$_2$S$_2$(SCH$_3$)]$^{2-}$ benchmark, together with the previous [Cu$_2$O$_2$]$^{2+}$ results~\cite{malone_ipie_2023}, indicate that MSD-AFQMC may be even a faithful alternative to exhaustive methods within the active space. Moreover, considering recent proposals for preparing AFQMC trials on quantum devices, this new feature may also attract interest from researchers working with quantum trials which typically involve numerous determinants~\cite{huggins_unbiasing_2022}.

We note that our implementation is still in its early stages and runs on a single GPU only. We expect it being straightforward to extend our Wick's theorem calculations to a multi-GPU setup, enabling the calculation of larger systems. Additionally, since evaluating properties other than ground-state energy is crucial, incorporating the auto-differentiation method into our GPU-MSD framework would be highly beneficial and of great interest to the community.


As demonstrated in our [Fe$_2$S$_2$(SCH$_3$)]$^{2-}$ benchmark, a substantial number of determinants may be required to accurately account for the dynamical correlation in these strongly correlated systems. Addressing dynamical correlation for larger iron-sulfur clusters could become intractable~\cite{sharma2014low,li2019electronic}. It would be interesting to explore whether a GPU implementation of MPS-AFQMC (AFQMC with matrix product state trials) could treat these systems more effectively~\cite{wouters2014projector,qin2020combination,jiang2024unbiasing}.

\section{Acknowledgements}

This concise preprint introduces a significant enhancement to the \texttt{ipie} package and complements an incoming \texttt{ipie} release paper. In the future, as more GPU-MSD features are developed, they will be incorporated into this document. The authors would like to thank Joonho Lee and Tong Jiang for their valuable discussions on the benchmarks. Additionally, YH and DL express gratitude to Jerry Chen from Nvidia for his suggestion on the use of cooperative groups.

\section{Author contributions}

YH and DL implemented the GPU code with assistance from JG. YH conducted the benchmark calculations with support from HP.

\appendix

\section{Implementation details}

For \texttt{get\textunderscore dets}, the output arrays have dimensions ($N_w$, $N_d$). Hence, we can rewrite the for loops in these axes as element-wise kernels that contain $N_w\cdot N_d$ threads in total. For \texttt{fill\textunderscore os}, we still only need element-wise kernels, the only difference from \texttt{get\textunderscore dets} is the output arrays have three dimensions: $N_w$, $N_d$ and $N_{\gamma}$. Therefore we have one more dimension to parallelize. In \texttt{build\textunderscore det\textunderscore matrix} and \texttt{build\textunderscore cofactor\textunderscore matrix} functions, we can further parallelize the $N_{exct}$ dimensions, which represents the order of excitation. For \texttt{get\textunderscore ss}, the $N_w$ and $N_d$ dimensions can still be parallelized but reduce operations are needed in the $N_{\gamma}$ dimension. So we add a serial for-loop inside each thread. The serial reduction operation can be accelerated by \texttt{cooperative groups} in CUDA.

The most tricky one is the \texttt{reduce\textunderscore CI} part. In this set of functions, we are updating the \texttt{walkers.CI}s with dimensions ($N_w$, $N_{acto}$, $N_{acte}$), where the last two dimensions denotes the number of active-space orbitals and electrons. However, we update them according to an array with dimensions ($N_w$, $N_d$). This mismatch results in a situation where different determinants from the trial contribute to the same \texttt{walkers.CI} element. In other words, we run into "collisions" when we try to add upon the values of the \texttt{walkers.CI} entries from each determinant. We use \texttt{atomicAdd} to resolve this problem, which performs the read, add and write operation in a single atomic operation without interference from other threads. 

For all of the n-fold excitation functions, a \texttt{cupy.linalg.det} operation is needed but can't be wrapped into a CUDA kernel. We dig this determinant calculation out of the for loops, compute them once for all indices of $N_{exct}$, $N_w$, $N_d$ and send the results into the kernels. In \texttt{build\textunderscore det\textunderscore matrix} and \texttt{build\textunderscore cofactor\textunderscore matrix} functions, we can further parallelize the $N_{exct}$ dimensions.



\bibliography{references}

\end{document}